\documentclass[11pt]{article}
\usepackage[letterpaper,margin=0.70in,top=0.66in,bottom=0.70in]{geometry}
\usepackage[T1]{fontenc}
\usepackage[utf8]{inputenc}
\usepackage{newtxtext,newtxmath}
\usepackage{microtype}
\usepackage{setspace}
\usepackage{amsmath,mathtools}
\usepackage{booktabs,tabularx,longtable,array,multirow,threeparttable}
\usepackage{siunitx}
\usepackage{enumitem}
\setlist{nosep,leftmargin=*}
\usepackage{graphicx}
\usepackage{caption,subcaption}
\usepackage{tikz}
\usetikzlibrary{arrows.meta,positioning,shapes.geometric,fit,backgrounds,calc}
\usepackage{pgfplots}
\pgfplotsset{compat=1.18}
\usepackage{xcolor}
\definecolor{ink}{RGB}{35,35,35}
\definecolor{midgray}{RGB}{100,100,100}
\definecolor{lightgray}{RGB}{225,225,225}
\usepackage{hyperref}
\hypersetup{colorlinks=true,linkcolor=ink,citecolor=ink,urlcolor=ink,pdfauthor={Ricardo Alonzo Fernandez Salguero},pdftitle={Testing Centralized and Polycentric Computational Planning}}
\usepackage[nameinlink,noabbrev]{cleveref}
\usepackage{fancyhdr}
\newcommand{\R}{\mathbb{R}}

\newcommand{\1}{\mathbf{1}}
\newcommand{\code}[1]{\texttt{#1}}

\title{\textbf{Testing Centralized and Polycentric Computational Planning:}\\[0.25em]
\large A Falsifiable Robust-Dominance Benchmark with Input--Output Networks, Agent-Based Markets, and Adversarial Stress Tests}
\author{Ricardo Alonzo Fernández Salguero}
\date{June 2026}

\begin{document}
\maketitle
\vspace{-1.2em}

\begin{abstract}
This paper presents and critically evaluates a reproducible synthetic benchmark for comparing a computational planner, an agent-based market, and an augmented ``meta-market'' under a common simulated economic environment. The benchmark combines a sparse Leontief input--output network, noisy demand observation, capacity constraints, heterogeneous firms, endogenous price adjustment, multi-component welfare evaluation, randomized welfare weights, structural shock families, an optimizing red-team adversary, an information-reporting experiment, and a sparse scaling test. In the reported run, the planner obtains lower mean welfare loss than both decentralized alternatives in all ten training worlds and all five nominal holdout worlds. Its mean loss is 0.1018 in training and 0.1100 in holdout, compared with 0.2679 and 0.2697 for the meta-market. The planner also remains superior in five adversarially searched cases, reduces essential unmet demand in a stylized fuel-logistics shock, and preserves very small numerical residuals in the Leontief calculations. A local reproduction recovers the substantive metrics exactly, with runtime variation attributable to the computing environment. An additional one-hundred-world diagnostic, using the unchanged model functions, likewise preserves the planner's internal dominance.

The central contribution, however, is methodological rather than ideological. The experiment demonstrates how a falsifiable comparative-institutions benchmark can be organized, but it does not establish the empirical superiority of planning. Several design choices favor the planner mechanically: the planner directly uses the production matrix while market firms do not; 40.7 percent of sectors in a representative world receive no assigned firm; rights violations for the planner are fixed at zero; several audit flags are asserted rather than tested; the nominal holdout set largely reuses training world families; and the claimed Groves/VCG-style incentive test is not a formal Groves mechanism. The results should therefore be interpreted as verification of a synthetic software architecture and as a research prototype for stronger empirical and institutional tests. The paper concludes with a concrete validation agenda based on common production technologies, structural holdouts, empirical calibration, global sensitivity analysis, uncertainty quantification, genuine mechanism-design tests, and independent red-team replication.
\end{abstract}


{\let\thefootnote\relax\footnote{DOI: \href{https://doi.org/10.5281/zenodo.20708560}{10.5281/zenodo.20708560}}}

\section{Introduction}

The feasibility of large-scale economic coordination has historically been framed as a contest between centralized calculation and decentralized price formation. That framing remains intellectually powerful, but it is too coarse for contemporary computational analysis. Modern economies contain hierarchical firms, public procurement systems, regulated networks, platform-mediated allocation, local governments, financial markets, logistics control towers, and algorithmic forecasting systems. Coordination is therefore neither purely centralized nor purely decentralized. It is implemented through layered institutions that collect information, compute plans, revise targets, form prices, and allocate authority at different scales. The relevant scientific problem is not whether an abstract planner or an abstract market is universally superior. It is whether explicit institutional architectures can be represented under common technological and informational constraints and then exposed to tests that can fail.

This paper analyzes a Python benchmark designed for that purpose. The benchmark creates synthetic economic ``worlds'' with 150 sectors, ten regions, seven social-need groups, 140 firms, sparse inter-industry dependencies, heterogeneous productivity, capacity limits, quality requirements, environmental intensity, informal supply, and ten dimensions of institutional attack. Three allocation systems operate in each world. The planner observes demand with error, combines that signal with local information, solves a Leontief system, applies capacity and reserve rules, and prioritizes essential sectors. The agentic market assigns firms to sectors, adjusts prices in response to excess demand, and updates output, quality, cash-mediated productivity, and reputation over ten periods. The meta-market adds stronger arbitrage and hedging terms. A separate evaluator maps outcomes into eleven loss components: essential unmet demand, total unmet demand, intergroup inequality, quality shortfall, corruption, rights violations, logistics failure, innovation shortfall, environmental burden, volatility, and administrative cost.

The computational design responds to an important weakness in many debates about planning: one institution is often judged by its ideal theorem while the other is judged by a historical failure. A simulation benchmark can instead place alternative decision systems inside one generated environment, disclose their information sets, define outcome metrics in advance, and search for counterexamples. This is the correct direction. Comparative institutional claims should be subjected to common-world evaluation, multiple welfare functions, distribution shifts, and adversarial scenarios rather than defended by verbal plausibility alone.

The benchmark's reported results are striking. Across ten training worlds, the planner's mean loss is 0.1018, while the agentic market and meta-market obtain 0.3072 and 0.2679. Across five nominal holdout worlds, the corresponding losses are 0.1100, 0.3127, and 0.2697. The planner wins under every sampled welfare vector. Five derivative-free adversarial searches also fail to reverse the ordering. In the stylized fuel-logistics case, essential unmet demand equals 0.0767 for the planner, 0.7806 for the market, and 0.6721 for the meta-market. The numerical input--output core is stable, with a spectral-radius estimate near 0.55 and residuals close to machine precision. These are meaningful internal results: the code runs, the accounting systems are coherent at the numerical level, and the benchmark produces deterministic, reproducible outputs under a fixed seed.

Yet strong internal results do not automatically imply strong institutional evidence. A benchmark can be fully reproducible and still embody asymmetric rules, insufficiently independent evaluation, weak holdouts, or outcome measures that align more closely with one system's design. This problem is especially acute when every final audit flag is true. In a genuinely falsifiable design, universal success should prompt an examination of whether the tests are difficult enough and whether failure modes are generated by the model rather than simply named in the output. The present paper therefore combines reconstruction with adversarial methodological review. It reports what the benchmark establishes, identifies what remains unestablished, and distinguishes numerical verification, internal validation, external validation, and policy transportability.

The study makes five contributions. First, it translates the code into a transparent mathematical and institutional description. The generated world, planner, market, evaluator, welfare aggregation, adversary, incentive experiment, historical-style shock, and scaling routine are each stated in formal terms. Second, it reproduces the reported run and shows that the core metrics are stable across computing environments. Third, it situates the benchmark in the literature on the socialist calculation debate, input--output economics, operations research, polycentric governance, agent-based computational economics, robust optimization, mechanism design, and simulation validation. Fourth, it performs supplementary diagnostics without changing the model's functions, including a one-hundred-world stress sample and a sector-coverage audit. Fifth, it develops a validation agenda capable of converting the current prototype into a more symmetric and empirically informative comparative-institutions laboratory.

The paper does not claim that decentralized markets are efficient by construction, nor that centralized planning is impossible by definition. Hayek's knowledge argument, Leontief's accounting system, Kantorovich's optimization methods, Ostrom's polycentric governance, and modern agent-based economics address different layers of the coordination problem. Prices can aggregate and communicate some decentralized information, but they do not automatically internalize rights, distribution, public goods, or environmental costs. Planning can coordinate interdependencies and prioritize essential needs, but it can also suffer from measurement error, local-knowledge loss, strategic reporting, bureaucratic delay, political capture, and model misspecification. A useful benchmark should allow each architecture to succeed or fail for empirically and institutionally interpretable reasons.

The remainder of the paper proceeds as follows. The next section reviews the state of the art. The methodology section reconstructs the benchmark and its mathematical structure. The results section reports the original run, local reproduction, and supplementary diagnostics. The discussion evaluates the meaning of the findings and the benchmark's structural asymmetries. The conclusion states the defensible contribution and the requirements for stronger future evidence.

\section{State of the Art}

\subsection{Economic calculation, dispersed knowledge, and institutional comparison}

The modern calculation debate began with the claim that common ownership of the means of production would eliminate the market prices required for rational comparison among alternative production plans. Mises argued that monetary calculation over capital goods depends on exchange relations generated by markets and private control of productive assets \cite{Mises1935}. Lange's response accepted the importance of scarcity prices but proposed that a central planning board could use trial-and-error adjustment rules, treating administered prices as signals and revising them when surpluses or shortages appeared \cite{Lange1936,Lange1937}. The debate therefore already contained an algorithmic structure: information is observed, a rule updates a vector of scarcity indicators, and decentralized production units react to those indicators.

Hayek shifted the emphasis from solving a known optimization problem to discovering and communicating knowledge that is dispersed, contextual, and often tacit \cite{Hayek1945}. In that view, the planner's challenge is not merely computational complexity. It is epistemic: the relevant data are not initially available as a complete table, and preferences, techniques, opportunities, and local constraints change while the system is operating. Bowles, Kirman, and Sethi reinterpret Hayek's market process as an algorithm rather than a proof of laissez-faire, emphasizing that the informational virtues of decentralized adjustment depend on institutions, competition, and the type of knowledge being processed \cite{Bowles2017}. This distinction matters for simulation. A credible benchmark must not provide the planner with the true state while requiring the market to discover it, or vice versa. It must explicitly model who knows what, when signals arrive, how costly they are, and how institutions update beliefs.

The benchmark examined here partially embraces this requirement. The planner does not observe true demand directly; it receives a multiplicatively noisy signal and a local-information correction. The market does not receive an exogenous loss value; its prices emerge from iterative excess-demand adjustment. These are improvements over simply assigning institution-specific performance scores. Nevertheless, the informational comparison remains incomplete because the planner has direct access to the input--output matrix, capacities, essential-sector labels, labor coefficients, and quality requirements, whereas the market agents do not use the same technological network. The experiment therefore represents different institutional algorithms but not yet a common discovery problem.

The calculation debate also concerns novelty and nonconvex change. Static allocation over known goods differs from entrepreneurial discovery of new products, production methods, and organizational forms. The code represents innovation only as an additive ``innovation miss'' term driven by an attack parameter. This captures vulnerability but not endogenous discovery. A stronger design would allow both planners and firms to search a changing technology space, make irreversible investments, learn from failures, and create new edges in the production network. Such extensions would connect the benchmark to evolutionary and complexity economics rather than confining it to a fixed-coefficient allocation problem \cite{Arthur2014}.

\subsection{Input--output analysis, linear programming, and production networks}

Leontief's input--output framework provides the benchmark's mathematical core \cite{Leontief1936,MillerBlair2009}. Let $A\in\R_+^{N\times N}$ denote the matrix of direct input coefficients and let $d\in\R_+^N$ denote final demand. Gross output $x$ satisfies
\begin{equation}
 x = Ax+d,
 \qquad
 x=(I-A)^{-1}d,
 \label{eq:leontief}
\end{equation}
when the inverse exists and is nonnegative. The Hawkins--Simon conditions characterize productive input--output systems; for a nonnegative matrix, the spectral condition $\rho(A)<1$ is sufficient for the Neumann series $(I-A)^{-1}=\sum_{k=0}^{\infty}A^k$ and for a nonnegative gross-output solution \cite{HawkinsSimon1949}. The code generates a sparse nonnegative matrix and rescales it toward a target spectral radius, thereby constructing a numerically productive system.

Kantorovich demonstrated that production planning could be formulated as constrained optimization with scarcity multipliers that play the role of accounting prices \cite{Kantorovich1960}. Linear programming and duality transformed planning from an informal balancing exercise into an explicit allocation problem over objectives and constraints \cite{Dantzig1963}. The benchmark computes a labor-value-like vector
\begin{equation}
 v=(I-A^{\mathsf T})^{-1}\ell,
 \label{eq:value}
\end{equation}
where $\ell$ is direct labor, and checks the identity $\ell^{\mathsf T}x=d^{\mathsf T}v$. This equality follows directly from \cref{eq:leontief,eq:value}. It is a useful numerical consistency test. Strictly speaking, however, it is not a complete Karush--Kuhn--Tucker verification because the code does not formulate and solve a constrained primal program with complementary slackness and dual feasibility. The label ``KKT duality'' in the console should therefore be read as a primal--dual accounting check, not as proof that all KKT conditions of a stated optimization problem have been tested.

Contemporary production-network research shows why the topology of $A$ matters. Sectoral shocks need not wash out when inter-industry linkages concentrate influence in central suppliers; network structure can amplify idiosyncratic disturbances into aggregate fluctuations \cite{Acemoglu2012}. A planning benchmark that includes a production network can therefore analyze systemic propagation, reserve allocation, and bottlenecks more realistically than a collection of independent sectors. The current code uses $A$ in the planner's gross-output calculation and in the labor-value and price diagnostics. It does not, however, require market firms to acquire intermediate inputs according to $A$. Consequently, the planner and market are not yet competing over the same physical production technology. The production network functions as a planner-side accounting system rather than as a shared constraint on all institutions.

\subsection{Cybernetics, computational planning, and polycentric governance}

The application of computation to planning has a longer history than current discussions of artificial intelligence suggest. Project Cybersyn in Chile attempted to combine near-real-time industrial reporting, statistical exception detection, cybernetic control, and organizational participation during the Allende government. Medina's historical analysis emphasizes both the project's technical ambition and its political, organizational, and material limitations \cite{Medina2011}. Cybersyn is important not because it proved that a national economy could be centrally optimized, but because it treated planning as a feedback architecture with distributed data collection, selective escalation, and human decision-making.

Cockshott and Cottrell later proposed that modern computation could support detailed planning in physical units, using input--output relations, labor accounting, and iterative balancing \cite{CockshottCottrell1993}. Their work directly challenges the view that the calculation problem is reducible to computational infeasibility. Yet the epistemic and institutional objections remain: data can be strategically distorted; objectives can conflict; innovation changes the feasible set; and computational tractability does not determine who controls the system or how rights are protected. The benchmark's inclusion of corruption, local knowledge, semantic attack, bureaucratic delay, and rights is therefore conceptually valuable. It recognizes that a computational planner is an institution embedded in incentives and power, not merely a matrix inversion.

Polycentric governance provides a more useful conceptual contrast than a binary opposition between state and market. Ostrom showed that complex systems can be governed by multiple, partially autonomous centers with overlapping authority, local monitoring, nested rules, and mechanisms for conflict resolution \cite{Ostrom2010}. Polycentricity is not synonymous with privatization. It describes an architecture in which local units retain knowledge and adaptive capacity while higher levels provide coordination, standards, redistribution, or conflict management. A future version of the benchmark could model regional planners, sector councils, firms, households, and central institutions as interacting nodes rather than treating the planner as a single algorithm. Such a design would better test the title's centralized/polycentric distinction.

\subsection{Agent-based computational economics and model validation}

Agent-based computational economics represents economies as evolving systems of heterogeneous, interacting agents rather than as solutions to a representative-agent equilibrium \cite{Tesfatsion2006,Epstein2006}. Farmer and Foley argued that agent-based models can complement conventional macroeconomic models by representing heterogeneity, network interaction, and out-of-equilibrium adjustment \cite{FarmerFoley2009}. The recent survey by Axtell and Farmer documents substantial applications in markets, industrial organization, labor, macroeconomics, development, finance, public policy, and environmental economics, while also emphasizing calibration, identification, computational demands, and validation challenges \cite{AxtellFarmer2025}.

The market module in the present benchmark belongs to this broad family. Firms are heterogeneous in productivity and cash, are assigned to sectors, and update supply in response to expected margins. Because they rely on simplified behavioral rules rather than solve a global optimization problem, their decision process is also consistent with Simon's conception of bounded rationality \cite{Simon1955}. Prices evolve through excess demand. Reputation changes stochastically, and quality evolves under innovation and corruption pressures. These features create endogenous price dispersion rather than imposing a precomputed market loss. The reported coefficient of variation of prices, 0.4357, confirms that the simulated market produces nontrivial dispersion.

Agent-based emergence alone does not validate a model. Fagiolo, Moneta, and Windrum describe empirical validation as a central weakness of agent-based economics and distinguish calibration, indirect calibration, history-friendly modeling, and statistical approaches \cite{Fagiolo2007}. Windrum, Fagiolo, and Moneta likewise stress the need to compare simulated and observed regularities rather than relying only on internal plausibility \cite{Windrum2007}. Grimm and coauthors' ODD protocol improves replicability by requiring a structured description of purpose, entities, state variables, scales, process scheduling, design concepts, initialization, input data, and submodels \cite{Grimm2020}. The methodology below follows that spirit, but the current benchmark remains a stylized synthetic model. It has not been calibrated to a national input--output table, firm distribution, household survey, price adjustment process, logistics network, corruption measure, or environmental account.

Verification and validation should be separated. Verification asks whether the implementation correctly solves the equations and follows the stated algorithm. Validation asks whether the model is an adequate representation for an intended purpose. Kleijnen's simulation methodology emphasizes code checking, sensitivity analysis, comparison with analytical results where possible, and empirical testing \cite{Kleijnen1995}. The benchmark performs several verification exercises: residual checks, reproducibility under a fixed seed, finite-number checks, and sparse scaling. Its external validation is intentionally absent. The code itself states that the output is ``still synthetic'' and not an implementation proof. This paper treats that disclaimer as a substantive boundary, not a minor caveat.

\subsection{Robust optimization, adversarial testing, and mechanism design}

Robust optimization studies decisions that remain feasible or perform acceptably under uncertain parameters. Bertsimas and Sim formalize the trade-off between nominal performance and protection against uncertainty through budgeted uncertainty sets \cite{BertsimasSim2004}; Ben-Tal, El Ghaoui, and Nemirovski provide a broader treatment of robust counterparts and adjustable decisions \cite{BenTal2009}. The benchmark does not solve a formal robust counterpart, but it adopts the robust-evaluation idea by sampling worlds, welfare vectors, and attack parameters and by searching for attacks that maximize the planner's relative disadvantage.

The adversary operates over ten normalized dimensions: data noise, cyber disruption, institutional capture, logistics attack, hidden-local-knowledge obscurity, preference drift, corruption pressure, innovation volatility, bureaucratic delay, and semantic attack. A derivative-free stochastic search perturbs the current best vector over eighteen iterations. This is stronger than testing a few hand-picked shocks, but it is not a global robustness certificate. Ten-dimensional black-box search with eighteen candidate steps can miss narrow failure regions. Robust dominance should therefore be stated as ``not falsified by the implemented search budget,'' not as ``cannot be broken.''

The benchmark also contains an incentive experiment. It compares truthful reports with multiplicative exaggeration under a quadratic score and a threshold-triggered audit penalty. Groves mechanisms create truthful incentives under specific quasi-linear environments by making each participant internalize the external effect of its report \cite{Groves1973}; Vickrey and Clarke provide related pivotal mechanisms \cite{Vickrey1961,Clarke1971}. Proper scoring rules elicit probabilistic beliefs when expected score is uniquely optimized by truthful reporting \cite{GneitingRaftery2007}. The implemented rule is neither a demonstrated Groves mechanism nor a standard strictly proper scoring rule over probability distributions. It is a stylized penalty system. Its strong result---a mean gain from lying of $-0.6708$---shows that the chosen penalties deter the chosen exaggeration pattern, but it does not prove dominant-strategy incentive compatibility for unilateral deviations, coalitions, alternative utility functions, or endogenous audit avoidance.

\subsection{Research gap}

The literature provides mature components: input--output accounting, optimization, decentralized knowledge, agent-based markets, polycentric governance, robust decision methods, mechanism design, and simulation validation. What is less common is an open comparative benchmark that combines these components under a common world and explicitly allows institutional claims to fail. The present code is valuable as a scaffold for such a benchmark. Its limitation is not that it is synthetic; all simulation begins with abstraction. Its limitation is that the current abstraction does not yet impose sufficiently symmetric technology, information, and falsification conditions to support a general causal claim about planning versus markets. The appropriate research objective is therefore to improve the benchmark's identification strategy, not merely to increase its scale or the number of passing audit flags.

\section{Methodology}

\subsection{Research design and computational provenance}

The analysis is based on the author-supplied Python program titled \emph{Centralized / Polycentric Computational Planning} and its complete console output. The program uses NumPy, pandas, SciPy sparse matrices, and concurrent execution. The fixed seed is 20,260,422. The full configuration uses $N=150$ sectors, $R=10$ regions, $G=7$ need groups, 140 firms, ten market iterations, ten training worlds, five nominal holdout worlds, 25 welfare draws, eighteen adversarial search steps, and a high-dimensional sparse test with 2,600 nodes. A fast smoke-test mode exists but was disabled in the reported run.

The supplied console output was reproduced by executing the source without modification. All deterministic substantive quantities reported below matched the supplied output. The original runtime was 6.9182 seconds; the local reproduction took 4.5278 seconds. The sparse fixed-point phase took 0.0081 seconds in the supplied run and 0.0058 seconds locally. These differences are expected because wall-clock time depends on processor, thread scheduling, numerical libraries, and background load. The local run also used the thread fallback in the parallel phase, as did the supplied output.

For the supplementary audit, the same functions were imported without altering their equations or parameter values. Two additional diagnostics were computed. First, firm-sector assignment was examined in the baseline world used by the market-emergence phase. Second, 100 additional worlds were generated across the five existing world families, with independent full-range attack vectors and 100 welfare draws. These diagnostics test the internal behavior of the unchanged model; they are not empirical validation.

The complete source code, computational models, simulations, and reproducibility materials are available in the associated Jupyter Notebook: \href{https://github.com/RAFS20/Planning/blob/main/CENTRALIZED_POLYCENTRIC_COMPUTATIONAL_PLANNING.ipynb}{\textbf{Access the code here.}}

\subsection{Overview of the benchmark architecture}

\begin{figure}[htbp]
\centering
\begin{tikzpicture}[
  node distance=8mm and 12mm,
  box/.style={draw=ink,rounded corners,align=center,minimum height=8mm,minimum width=27mm,fill=lightgray!45,font=\small},
  proc/.style={draw=ink,align=center,minimum height=10mm,minimum width=31mm,font=\small},
  arrow/.style={-{Latex[length=2mm]},thick,draw=ink}
]
\node[box] (world) {World generator\\$A,d,c,\ell,q,r,g,e,a$};
\node[proc,below left=15mm and 29mm of world] (planner) {Planner\\noisy demand + IO solve};
\node[proc,below=15mm of world] (market) {Agentic market\\firms + price adjustment};
\node[proc,below right=15mm and 29mm of world] (meta) {Meta-market\\arbitrage + hedging};
\node[box,below=18mm of market] (judge) {External judge\\11 loss components};
\node[box,below=12mm of judge] (weights) {Welfare draws\\Dirichlet weights};
\node[box,left=16mm of weights] (adv) {Adversary\\10 attack dimensions};
\node[box,right=16mm of weights] (audit) {Audit board\\44 Boolean checks};
\draw[arrow] (world) -- (planner);
\draw[arrow] (world) -- (market);
\draw[arrow] (world) -- (meta);
\draw[arrow] (planner) -- (judge);
\draw[arrow] (market) -- (judge);
\draw[arrow] (meta) -- (judge);
\draw[arrow] (judge) -- (weights);
\draw[arrow] (adv) -- (world);
\draw[arrow] (weights) -- (audit);
\end{tikzpicture}
\caption{Computational architecture of the benchmark. The diagram represents the intended separation between world, institutions, judge, welfare aggregation, and adversarial evaluation. The degree of substantive independence is assessed in the discussion.}
\label{fig:architecture}
\end{figure}
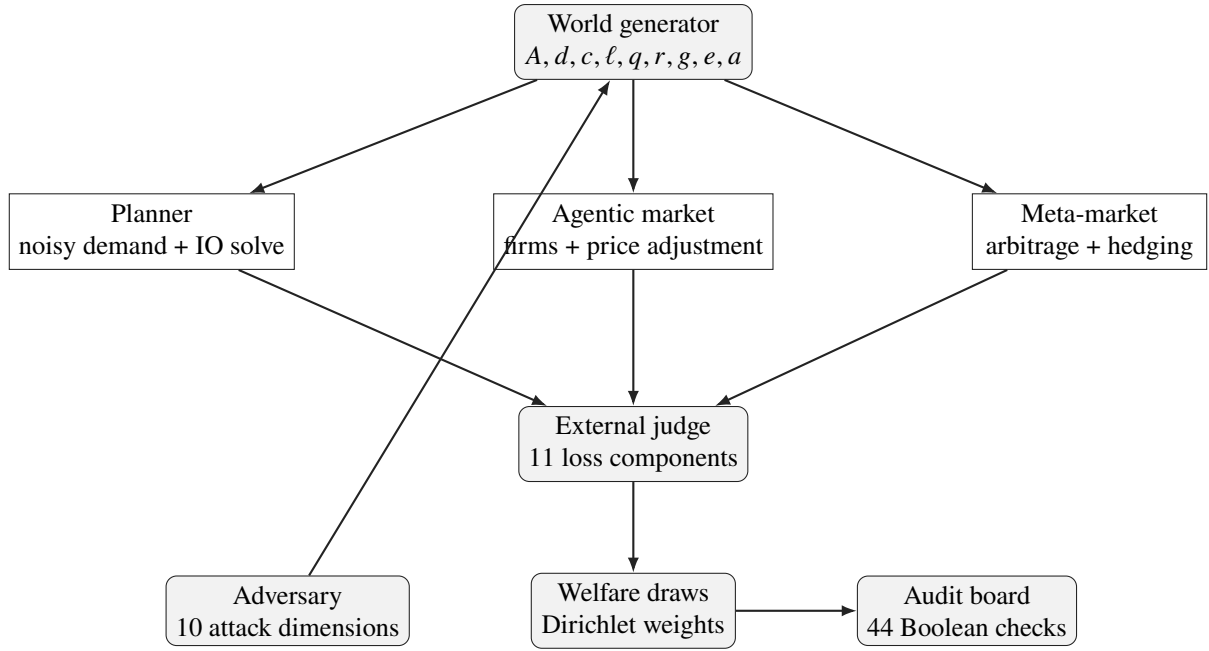

\Cref{fig:architecture} summarizes the information flow. A world generator creates the physical and institutional state. The planner and two market architectures receive different operational signals and produce outcome vectors. An external judge maps those outcomes and the hidden world state into eleven loss components. Welfare weights aggregate the components. Separate modules test mathematical consistency, ensemble performance, adversarial robustness, reporting incentives, a historical-style shock, and computational scaling.

\subsection{Synthetic world generation}

Each world is a tuple
\begin{equation}
 \mathcal{W}=(A,\rho,\ell,d,c,b,q,s,G,E,F,\pi,h,\iota,z,\eta,a,\tau),
\end{equation}
where $A$ is the input--output matrix; $\rho$ is its estimated spectral radius; $\ell$ is direct labor; $d$ is demand; $c$ is capacity; $b$ is base cost; $q$ is required quality; $s$ maps sectors to regions; $G$ contains sector-by-group need shares; $E$ marks essential sectors; $F$ maps firms to sectors; $\pi$ is firm productivity; $h$ is hidden local information; $\iota$ is regional informality; $z$ contains logistics coordinates; $\eta$ is environmental intensity; $a\in[0,1]^{10}$ is the attack vector; and $\tau$ is the world type.

The sparse input--output matrix is generated with nonnegative gamma-distributed entries and zero diagonal. If $A_0$ is the raw matrix and $\widehat\rho(A_0)$ is the power-iteration estimate, the program sets
\begin{equation}
 A=\frac{\rho^{\star}}{\widehat\rho(A_0)+10^{-12}}A_0,
\end{equation}
with target $\rho^{\star}=0.55$ in ordinary worlds and $0.82$ in the \code{near\_unstable} family. The density is 0.016. Demand is Pareto-distributed and multiplied by 900; labor and environmental intensity are lognormal; capacity equals demand times a uniform factor between 1.08 and 1.45; and base cost is a labor-scaled random variable. Quality needs lie between 0.68 and 0.96. Group need shares are Dirichlet draws. At least eighteen sectors are designated essential.

Five world families alter these primitives:
\begin{enumerate}[label=(\roman*)]
\item \code{baseline}: the default distributions;
\item \code{heavy\_tail}: approximately one-twelfth of sectors receive additional demand multipliers between 2.5 and 5;
\item \code{supply\_crunch}: capacity is multiplied by a random factor between 0.72 and 0.95;
\item \code{near\_unstable}: the target input--output radius rises from 0.55 to 0.82;
\item \code{historical\_fuel\_logistics}: approximately one-tenth of sectors receive a demand multiplier of 1.55.
\end{enumerate}

This construction generates heterogeneity and stress but is not calibrated to observed sectoral distributions. The random matrix is productive by construction after scaling. Demand, capacity, and costs are linked through formulas selected by the modeler rather than estimated from data. Accordingly, the worlds are best interpreted as controlled test fixtures.

\subsection{Planner policy}

The planner receives the full structural world and a noisy observation of demand. Let $a_k$ denote the $k$th attack component, ordered as data noise, cyber disruption, capture, logistics attack, hidden-local obscurity, preference drift, corruption pressure, innovation volatility, bureaucratic delay, and semantic attack. Observed demand is
\begin{equation}
 d_i^{\mathrm{obs}}=d_i\exp(\varepsilon_i),
 \qquad
 \varepsilon_i\sim\mathcal{N}\!\left(0,\,0.04+0.22a_1\right).
\end{equation}
The signal is winsorized at the third and ninety-seventh percentiles. It is then multiplied by a local-information correction based on the regional hidden-information matrix plus Gaussian error. The resulting $\widehat d$ is strictly positive.

The planner solves the Leontief system
\begin{equation}
 x^{P}=(I-A)^{-1}\widehat d.
\end{equation}
If sparse direct solution fails, an eighty-step fixed-point iteration is used. Effective capacity is
\begin{equation}
 \widetilde c_i=c_i(1-0.38a_4)(1-0.12a_2).
\end{equation}
Essential sectors receive reserve multipliers of $1+0.20+0.20a_4+0.10a_6$; nonessential sectors receive one. Planned supply is initially $\min(x_i^P r_i,\widetilde c_i)$. Essential supply is then raised toward at least 94 percent of final demand when capacity permits. This rule directly aligns the planner with the judge's essential-unmet-demand component.

Quality investment is based on a shadow-cost proxy
\begin{equation}
 \widetilde\ell=\ell+A^{\mathsf T}\ell,
\end{equation}
which is normalized and combined with an essential-sector premium. Delivered quality is a bounded function of required quality, the investment proxy, and cyber disruption. Corruption, logistics failure, innovation miss, volatility, and administrative cost are computed from deterministic piecewise-linear functions of the attack vector. Planner rights violations are assigned the constant value zero.

The planner is thus neither omniscient nor an optimizer of the final welfare function. It has demand noise and does not directly minimize the judge's weighted loss. Nevertheless, it has comprehensive structural access and several rules that target judge components. It uses $A$, capacities, essential labels, labor, required quality, and hidden local information. The significance of that information advantage is examined later.

\subsection{Agentic market and meta-market}

The market initializes sectoral prices from base cost and a random markup. Each of 140 firms is assigned to one of 150 sectors. Firm productivity is lognormal, cash is lognormal, and reputation begins at one. Over $T=10$ iterations, households generate demand
\begin{equation}
 d_{it}^{M}=d_i\exp[-0.42(p_{it}/\widetilde p_t-1)]\exp(\nu_{it}),
\end{equation}
where $\widetilde p_t$ is the median price and the noise variance increases with preference drift. A firm's expected margin is price minus an attack-adjusted base cost. Its output contribution is proportional to productivity, cash, reputation, and a logistic transformation of expected margin. Cyber and logistics attacks reduce output.

Sectoral supply adds firm output, an informal-supply term, and an exogenous hedging term. For the ordinary market, the hedge coefficient is 0.10; for the meta-market it is 0.22. Supply is capped by attack-adjusted capacity. Prices update according to
\begin{equation}
 p_{i,t+1}=p_{it}\exp\left[\operatorname{clip}\left(0.18(1-\gamma)\frac{d_{it}^{M}-y_{it}}{y_{it}+0.05d_i+10^{-9}},-0.22,0.25\right)\right],
 \label{eq:priceupdate}
\end{equation}
where $\gamma=0.22$ in the ordinary market and $0.34$ in the meta-market. Quality evolves through a small innovation gain, stochastic noise, and penalties from corruption and cyber disruption. Reputation declines for randomly selected ``bad'' firms and gradually rises otherwise. The meta-market receives stronger arbitrage smoothing and hedging but a slightly lower innovation gain.

The model calls these markets ``emergent'' because prices are generated through repeated agent interactions rather than being assigned from the welfare function. That characterization is reasonable in a limited computational sense. However, the firms do not buy intermediate inputs through $A$, do not enter unserved sectors, do not invest in capacity, and do not reallocate capital across sectors except through the fixed formulas. Market supply is therefore an adaptive partial-equilibrium heuristic, not a general competitive equilibrium or a fully developed agent-based economy.

\subsection{External welfare judge}

For each institutional output, the judge computes nonnegative supply $y$, bounded quality, and unmet demand
\begin{equation}
 u_i=\frac{\max(d_i-y_i,0)}{d_i+10^{-9}}.
\end{equation}
Essential unmet demand is the mean of $u_i$ over essential sectors; total unmet demand is the mean over all sectors. Group satisfaction is the need-share-weighted delivered fraction, and inequality is its coefficient of variation. Quality loss is the mean positive difference between required and delivered quality.

The remaining components are corruption rent, rights violations, logistics failure, innovation miss, normalized environmental intensity of delivered supply, price volatility, and administrative cost. Let
\begin{equation}
 c=(c_1,\ldots,c_{11})^{\mathsf T}
\end{equation}
collect these components in the order stated above. For welfare vector $w\in\Delta^{10}$, total loss is
\begin{equation}
 L(w)=w^{\mathsf T}c.
 \label{eq:loss}
\end{equation}
Lower values are preferred.

Weights are sampled from a symmetric Dirichlet distribution. The code then replaces the essential-unmet and rights weights by at least 0.12 and renormalizes. After renormalization, their final values can fall below 0.12 because the lower bound is applied before the vector is divided by its new sum. The procedure nonetheless increases their expected importance relative to unconstrained Dirichlet draws. Twenty-five weight vectors are used in the main benchmark.

The judge is implemented as a separate function, which is good software modularity. Substantive independence is more demanding. Some institutional outputs enter directly as precomputed scalar penalties, and the planner's rights output is fixed at zero. In addition, supply is compared to final demand even though the planner computes gross output from the Leontief inverse. This mixes accounting concepts: gross output includes intermediate requirements, while the judge treats the resulting planned quantity as directly deliverable against final demand. The discussion considers the potential bias.

\subsection{Core mathematical validation}

The first benchmark phase generates a baseline world and computes gross output, embodied labor values, and a price vector. Values satisfy \cref{eq:value}. The price vector is obtained from
\begin{equation}
 p=\left[I-(1+r)A^{\mathsf T}\right]^{-1}(1+r)0.55\ell,
 \qquad
 r=0.35\left(\frac{1}{\rho}-1\right),
\end{equation}
and is rescaled to have the same sum as $v$. The tests require $\rho<1$, small residuals in the value and plan equations, a small primal--dual accounting gap, positive finite solutions, and a correlation above 0.85 between $v$ and $p$.

This phase is a strong verification test for sparse linear algebra. It confirms that the generated matrix produces well-conditioned solutions at the tested scale. It does not establish that labor values predict observed prices, because both vectors are constructed from the same synthetic matrix and labor coefficients.

\subsection{Ensemble and nominal holdout evaluation}

Training worlds cycle through baseline, heavy-tail, supply-crunch, and near-unstable families. Their attack vectors are uniform draws on $[0,0.42]^{10}$. Holdout worlds cycle through historical-fuel-logistics, heavy-tail, supply-crunch, baseline, and near-unstable families, with attacks on $[0,0.58]^{10}$. The same 25 welfare vectors are applied to all worlds. For each institution, the benchmark records mean loss, planner win rates, and the maximum ratio of planner loss to comparator loss across welfare vectors.

The term ``holdout'' should be interpreted carefully. Seeds and attack magnitudes are held out, but four of the five holdout world families also appear in training. Only the historical-fuel-logistics label is absent from the training family list, and its generator differs from baseline mainly by a demand multiplier on a subset of sectors. This is an out-of-sample random-seed test within the same data-generating framework, not a structural holdout under a new model class.

\subsection{Optimizing adversary}

For each of five world types, the adversary searches $a\in[0,1]^{10}$. Its objective is
\begin{equation}
 S(a)=\overline L_P(a)-\overline L_X(a)
 +0.35\max\left(0,\max_w\frac{L_P(w,a)}{L_X(w,a)}-1\right)
 +5R_P(a),
\end{equation}
where $X$ denotes the meta-market and $R_P$ is planner rights loss. The search initializes a random vector below 0.65, perturbs the current vector with Gaussian noise, clips to the unit cube, and applies a small simulated-annealing-like acceptance probability. The perturbation scale starts at 0.32 and decays by 1.5 percent per step.

Because planner rights are identically zero, the final term cannot identify a planner rights failure. The adversary mainly maximizes relative welfare loss. Eighteen steps across ten dimensions provide a nontrivial stress test but not exhaustive optimization. No convergence diagnostics, repeated adversarial restarts, surrogate model, or global upper bound is computed.

\subsection{Information-reporting experiment}

The incentive phase generates 1,500 true needs from a lognormal distribution. A noisy signal has lognormal multiplicative error. Truthful reports equal true needs; false reports multiply needs by independent factors between 1.15 and 2.4. For a vector of reports $r$, normalized allocation is
\begin{equation}
 A_i(r)=\frac{r_i}{\overline r+10^{-12}}.
\end{equation}
Utility is defined as
\begin{equation}
 U_i(r)=A_i(r)
 -0.42\left(\frac{r_i-s_i}{s_i+10^{-9}}\right)^2
 -0.38\,\mathbf{1}\left\{\left|\frac{r_i-s_i}{s_i+10^{-9}}\right|>0.24\right\}A_i(r),
 \label{eq:incentive}
\end{equation}
where $s_i$ is the audit signal. The experiment reports the mean and ninety-fifth percentile of $U_i(r^{\mathrm{lie}})-U_i(r^{\mathrm{truth}})$ and the proportion of lies detected by the threshold.

The pass rule requires mean gain from lying below 0.01, ninety-fifth percentile below 0.08, and detection above 0.75. This is a useful behavioral stress test of a penalty schedule. It is not a formal proof of incentive compatibility because all agents change reports simultaneously, the signal is treated as an audit truth proxy, utilities are imposed rather than derived, and deviations outside the specified exaggeration family are not tested.

\subsection{Historical-style scenario and scaling test}

The historical-style phase uses the \code{historical\_fuel\_logistics} generator and the fixed attack vector
\begin{equation}
 (0.38,0.22,0.28,0.55,0.20,0.35,0.26,0.24,0.30,0.18).
\end{equation}
It is described as a fuel, logistics, import, and currency-pressure shock. No empirical time series or historical parameter estimate enters the generator. The case should therefore be called a stylized reconstruction rather than a calibrated historical simulation.

The scaling phase creates a sparse 2,600-by-2,600 matrix with density 0.0005, rescales it, and performs ninety fixed-point iterations for values and output. It then evaluates eight independent baseline worlds using process-based parallelism when possible and thread-based fallback otherwise. This tests sparse matrix operations and embarrassingly parallel evaluation. It does not run the full 2,600-sector planner-market-adversary benchmark; the high-dimensional test is limited to the linear core.

\subsection{Evaluation criteria}

The paper classifies evidence into four levels:
\begin{enumerate}[label=\textbf{Level \arabic*:}]
\item \textbf{Numerical verification}: equations are solved with small residuals, outputs are finite, and fixed seeds reproduce results.
\item \textbf{Internal comparative validity}: under the encoded world and rules, institutional comparisons are computed consistently and can in principle reverse.
\item \textbf{External empirical validity}: parameters, behavioral rules, and shock distributions match observed economic data with quantified uncertainty.
\item \textbf{Policy transportability}: conclusions remain valid under institutional implementation, political constraints, strategic adaptation, legal rights, and distribution shift.
\end{enumerate}
The current benchmark provides substantial Level 1 evidence and partial Level 2 evidence. It does not yet provide Levels 3 or 4.

\section{Results}

\subsection{Numerical verification of the input--output core}

\Cref{tab:core} reports the core diagnostics. The estimated spectral radius is 0.5500000001, safely below one. Maximum absolute residuals are approximately $4.9\times10^{-15}$ for the value equation and $2.5\times10^{-12}$ for the gross-output equation. The accounting gap between direct-labor cost of gross output and value-weighted final demand is $1.67\times10^{-16}$ relative to the primal quantity. All values, outputs, and prices are positive and finite. The synthetic price and embodied-labor vectors have correlation 0.9643.

\begin{table}[htbp]
\centering
\caption{Core mathematical verification}
\label{tab:core}
\begin{threeparttable}
\begin{tabular}{lrr}
\toprule
Diagnostic & Reported value & Pass threshold \\
\midrule
Estimated spectral radius $\widehat\rho(A)$ & 0.5500000001 & $<1$ \\
Maximum value-equation residual & $4.88498\times10^{-15}$ & $<10^{-8}$ \\
Maximum plan-equation residual & $2.50111\times10^{-12}$ & $<10^{-7}$ \\
Relative primal--dual accounting gap & $1.67316\times10^{-16}$ & $<10^{-8}$ \\
Correlation between constructed values and prices & 0.964272 & $>0.85$ \\
Positive values, plan, and finite numerics & True & True \\
\bottomrule
\end{tabular}
\begin{tablenotes}[flushleft]\footnotesize
\item The values are generated from the same synthetic technology and labor coefficients. The correlation is therefore a consistency result, not an empirical price-value test.
\end{tablenotes}
\end{threeparttable}
\end{table}

The numerical evidence is strong for the stated linear systems. It verifies the sparse solve and the algebraic identities. It also confirms that the near-machine-precision residuals are not artifacts of the supplied console transcript: the local reproduction recovered the same substantive quantities. This is the benchmark's clearest result because it depends on explicit mathematical equations and direct residual checks.

\subsection{Market emergence and planner information}

The market-emergence phase reports a price coefficient of variation of 0.4357. Prices therefore do not collapse to a constant or remain at their initial values. The planner's demand-estimation mean absolute percentage error is 0.03274, above the code's 0.025 threshold for ``limited information.'' This demonstrates that the planner receives a noisy demand signal rather than exact demand.

These checks are necessary but not sufficient for institutional separation. Price dispersion establishes endogenous movement, not market efficiency. Likewise, a 3.27 percent demand error establishes imperfect observation, but the planner still receives all other structural arrays. Several separation flags are set to true as literal constants: evaluator separation, world-planner-market-judge separation, absence of hard-coded market loss, and the ability of the benchmark to fail. The software architecture does separate functions, but those Boolean outputs do not independently test the economic symmetry of the modules.

\subsection{Training and nominal holdout performance}

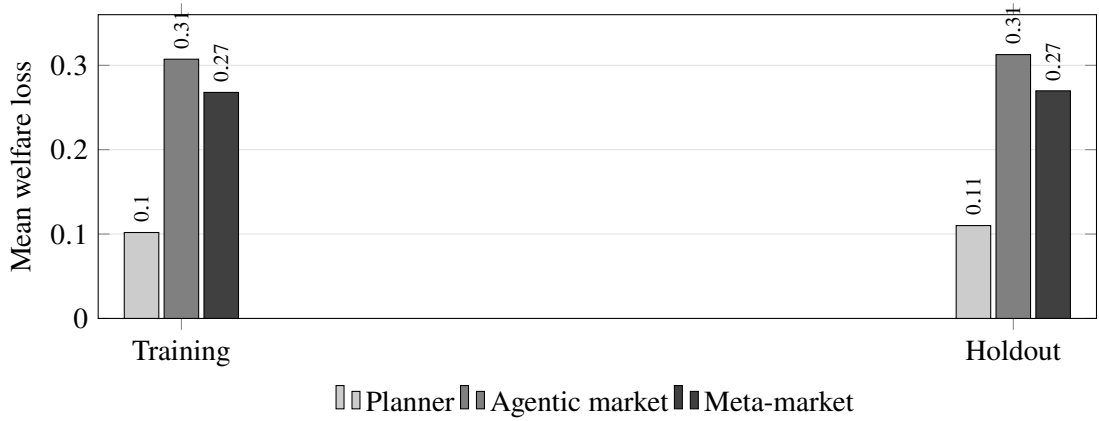
\begin{figure}[htbp]
\centering
\begin{tikzpicture}
\begin{axis}[
 ybar,
 width=0.82\linewidth,
 height=5.6cm,
 ylabel={Mean welfare loss},
 symbolic x coords={Training,Holdout},
 xtick=data,
 ymin=0,ymax=0.36,
 bar width=13pt,
 legend style={at={(0.5,-0.20)},anchor=north,legend columns=3,draw=none},
 nodes near coords,
 nodes near coords style={font=\scriptsize,rotate=90,anchor=west},
 ymajorgrids=true,
 grid style={lightgray}
]
\addplot[fill=black!20,draw=black] coordinates {(Training,0.1018396878) (Holdout,0.1100449255)};
\addplot[fill=black!50,draw=black] coordinates {(Training,0.3072323121) (Holdout,0.3126918638)};
\addplot[fill=black!75,draw=black] coordinates {(Training,0.2678788670) (Holdout,0.2697330823)};
\legend{Planner,Agentic market,Meta-market}
\end{axis}
\end{tikzpicture}
\caption{Mean welfare loss in the supplied benchmark. Lower values indicate better performance under the encoded judge and welfare draws.}
\label{fig:meanloss}
\end{figure}

\Cref{tab:aggregate} and \cref{fig:meanloss} show the aggregate loss results. The planner's mean loss is approximately one-third of the agentic market's loss and below half of the meta-market's loss in both samples. It wins against the meta-market under all sampled welfare weights in every world. The largest observed planner-to-meta-market loss ratio is 0.7491 in training and 0.7845 in holdout, both below one.

\begin{table}[htbp]
\centering
\caption{Aggregate institutional performance}
\label{tab:aggregate}
\resizebox{\linewidth}{!}{%
\begin{tabular}{lrrrrrr}
\toprule
Sample & Worlds & Planner & Agentic market & Meta-market & Planner win rate vs. meta & Maximum worst ratio \\
\midrule
Training & 10 & 0.101840 & 0.307232 & 0.267879 & 1.000 & 0.749067 \\
Nominal holdout & 5 & 0.110045 & 0.312692 & 0.269733 & 1.000 & 0.784521 \\
\bottomrule
\end{tabular}}
\end{table}

The world-level results in \cref{tab:worlds} show that the aggregate finding is not driven by a single case. Planner mean loss ranges from 0.0943 to 0.1200 in training and from 0.0961 to 0.1370 in holdout. The largest holdout planner loss occurs in the supply-crunch world. The meta-market is consistently better than the ordinary agentic market, reflecting its larger hedging term and stronger price smoothing, but it remains far behind the planner.

\begin{table}[htbp]
\centering
\caption{World-level mean loss and robustness ratio}
\label{tab:worlds}
\scriptsize
\begin{tabular}{clrrrrr}
\toprule
Set & World type & Planner & Market & Meta & $P$ win rate vs. $X$ & Worst $P/X$ \\
\midrule
Train & Baseline & 0.096084 & 0.311320 & 0.273396 & 1.00 & 0.685001 \\
Train & Heavy tail & 0.094807 & 0.303165 & 0.263460 & 1.00 & 0.723485 \\
Train & Supply crunch & 0.120049 & 0.308658 & 0.269555 & 1.00 & 0.737116 \\
Train & Near unstable & 0.099859 & 0.301290 & 0.266805 & 1.00 & 0.724374 \\
Train & Baseline & 0.099214 & 0.307216 & 0.266672 & 1.00 & 0.716955 \\
Train & Heavy tail & 0.106415 & 0.308137 & 0.266213 & 1.00 & 0.731198 \\
Train & Supply crunch & 0.108672 & 0.303260 & 0.266001 & 1.00 & 0.749067 \\
Train & Near unstable & 0.095192 & 0.305804 & 0.267508 & 1.00 & 0.695180 \\
Train & Baseline & 0.103852 & 0.308759 & 0.270445 & 1.00 & 0.726669 \\
Train & Heavy tail & 0.094252 & 0.314714 & 0.268734 & 1.00 & 0.681061 \\
\midrule
Holdout & Historical fuel/logistics & 0.108823 & 0.309611 & 0.269604 & 1.00 & 0.728772 \\
Holdout & Heavy tail & 0.096129 & 0.316253 & 0.269478 & 1.00 & 0.686802 \\
Holdout & Supply crunch & 0.137040 & 0.312885 & 0.268027 & 1.00 & 0.784521 \\
Holdout & Baseline & 0.108434 & 0.308139 & 0.268536 & 1.00 & 0.719446 \\
Holdout & Near unstable & 0.099799 & 0.316571 & 0.273020 & 1.00 & 0.719955 \\
\bottomrule
\end{tabular}
\end{table}

The stability of the ordering is internally impressive. It is also diagnostically important. When one architecture wins in every world and every welfare draw, the result can mean either that the institution is genuinely robust within the model or that the comparison contains a persistent structural advantage. The supplementary audits below provide evidence for both interpretations: the planner is robust under unchanged code, but the market module is structurally supply-constrained in ways not shared by the planner.

\subsection{Adversarial search}

The adversary searches for high attack vectors that make the planner lose to the meta-market. \Cref{tab:adversary} reports the best attacks found. All objective scores remain negative, meaning that planner mean loss stays below meta-market mean loss even after adding the ratio penalty. The supply-crunch case comes closest to reversal: planner loss is 0.23674, meta-market loss is 0.31810, and the maximum welfare-specific ratio reaches 0.9568. Planner rights remain zero in every case.

\begin{table}[htbp]
\centering
\caption{Best attacks found by the derivative-free adversary}
\label{tab:adversary}
\small
\begin{tabular}{lrrrr}
\toprule
World type & Adversary score & Planner loss & Meta-market loss & Worst $P/X$ \\
\midrule
Baseline & -0.093498 & 0.18881 & 0.28230 & 0.8526 \\
Heavy tail & -0.097077 & 0.15943 & 0.25651 & 0.8315 \\
Supply crunch & -0.081363 & 0.23674 & 0.31810 & 0.9568 \\
Near unstable & -0.122163 & 0.16259 & 0.28475 & 0.8803 \\
Historical fuel/logistics & -0.104504 & 0.15255 & 0.25705 & 0.8265 \\
\bottomrule
\end{tabular}
\end{table}

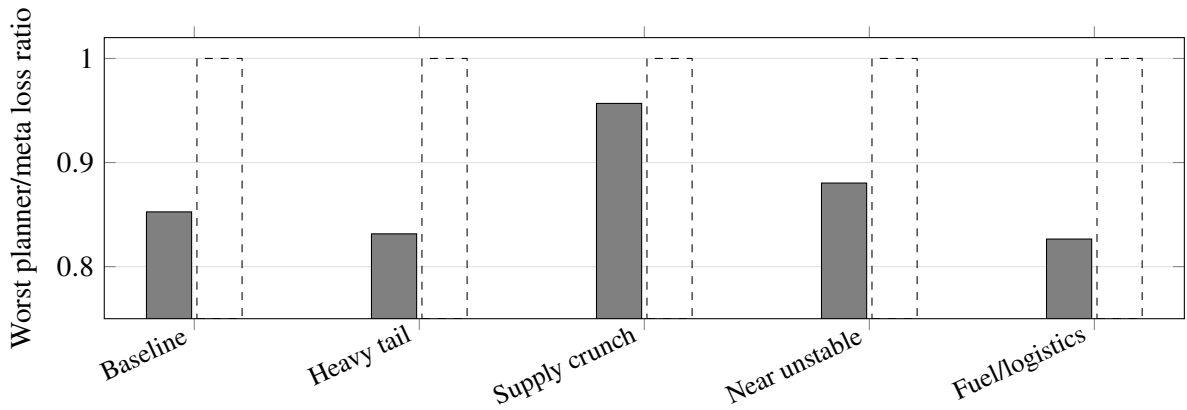
\begin{figure}[htbp]
\centering
\begin{tikzpicture}
\begin{axis}[
 width=0.88\linewidth,height=5.3cm,
 ybar,
 ymin=0.75,ymax=1.02,
 ylabel={Worst planner/meta loss ratio},
 symbolic x coords={Baseline,Heavy tail,Supply crunch,Near unstable,Fuel/logistics},
 xtick=data,
 x tick label style={rotate=25,anchor=east,font=\small},
 bar width=17pt,
 ymajorgrids=true,grid style={lightgray}
]
\addplot[fill=black!50,draw=black] coordinates {(Baseline,0.8526) (Heavy tail,0.8315) (Supply crunch,0.9568) (Near unstable,0.8803) (Fuel/logistics,0.8265)};
\addplot[black,dashed,mark=none] coordinates {(Baseline,1) (Heavy tail,1) (Supply crunch,1) (Near unstable,1) (Fuel/logistics,1)};
\end{axis}
\end{tikzpicture}
\caption{Maximum planner-to-meta-market loss ratio found in each adversarial case. A ratio above one would indicate a welfare draw under which the planner loses.}
\label{fig:adversary}
\end{figure}

Within the implemented search, the adversary does not overturn the result. The appropriate statistical wording is that no counterexample was found under five fixed seeds, one initialization per world type, and eighteen search steps. The output flag \code{ADAPTIVE\_RED\_TEAM\_CANNOT\_BREAK\_SYSTEM} is stronger than the evidence. A more defensible label would be \code{NO\_REVERSAL\_FOUND\_UNDER\_SEARCH\_BUDGET}.

\subsection{Information-reporting experiment}

The reporting experiment produces a mean utility difference between lying and truth-telling of $-0.6708$. The ninety-fifth percentile is $-0.2463$, so even relatively favorable lying outcomes remain below truthful utility. The threshold detects 91.47 percent of the generated exaggerations. All four audit checks in this phase pass.

These results show that the imposed quadratic discrepancy cost and audit penalty are powerful relative to the allocation gain from exaggeration. They do not reveal a subtle emergent incentive property; the penalty function is designed to grow rapidly with report-signal distance. The result is useful as a unit test for the penalty rule. It should not be presented as evidence that a Groves mechanism has been implemented, because no counterfactual externality payment is calculated and no dominant-strategy theorem is tested.

\subsection{Stylized fuel-logistics shock}

\begin{figure}[htbp]
\centering
\begin{tikzpicture}
\begin{axis}[
 ybar,width=0.70\linewidth,height=5.2cm,
 ylabel={Essential unmet demand},
 symbolic x coords={Planner,Agentic market,Meta-market},
 xtick=data,ymin=0,ymax=0.9,
 bar width=25pt,
 nodes near coords,
 nodes near coords style={font=\small},
 ymajorgrids=true,grid style={lightgray}
]
\addplot[fill=black!45,draw=black] coordinates {(Planner,0.0766697) (Agentic market,0.7806198) (Meta-market,0.6720672)};
\end{axis}
\end{tikzpicture}
\caption{Essential unmet demand in the stylized historical fuel-logistics shock.}
\label{fig:essential}
\end{figure}
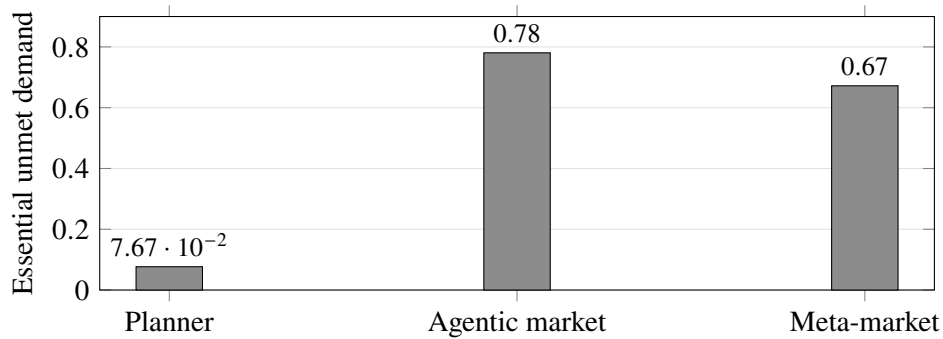

The historical-style case strongly favors the planner. Mean loss is 0.1060 for the planner, 0.3030 for the agentic market, and 0.2569 for the meta-market. Essential unmet demand is 0.0767 for the planner, 0.7806 for the market, and 0.6721 for the meta-market. \Cref{fig:essential} displays the difference.

The result is consistent with the planner's design: essential sectors are explicitly raised toward 94 percent of demand when capacity permits. The market has no parallel rationing or essential-goods institution. Its ordinary firm allocation and price adjustment operate over ten periods, while 40 percent or more of sectors can lack a firm. The historical-style result therefore demonstrates the effectiveness of an essential-priority rule relative to the particular market heuristic. It does not identify the effect of planning as such, because a decentralized system with strategic reserves, emergency procurement, price caps plus rationing, public logistics, entry, or targeted transfers is not modeled.

\subsection{Scaling and parallel execution}

The 2,600-node sparse phase reports an estimated radius of 0.4785, maximum residuals of $3.11\times10^{-15}$ and $1.82\times10^{-12}$, and 0.0081 seconds for ninety fixed-point iterations in the supplied environment. Eight independent baseline world evaluations are completed through a thread pool, with planner dominance in all eight. The phase confirms that the sparse linear core is computationally light at this dimension and that independent worlds can be evaluated in parallel.

The result should not be interpreted as a 2,600-sector full institutional simulation. The high-dimensional object includes the input--output fixed point but not 2,600-sector market agents, adversarial optimization, local governance, data ingestion, or dynamic replanning. Real operational complexity would be dominated by data quality, model updating, distributed systems, security, legal review, and human decision processes rather than by the matrix-vector products alone.

\subsection{Reproduction and supplementary diagnostics}

The local reproduction recovered the same spectral radius, residuals, correlations, institutional mean losses, adversarial outcomes, incentive statistics, historical-style results, and 44 passing audit flags. This establishes deterministic reproducibility for the supplied source and seed. The runtime difference does not affect the conclusions.

A supplementary one-hundred-world diagnostic used the unchanged generators and policy functions, cycled through all five existing world families, drew attack vectors on the full unit cube, and evaluated 100 welfare vectors. Planner mean loss was 0.1332, compared with 0.3326 for the agentic market and 0.2870 for the meta-market. The planner beat both comparators in all 100 worlds, and the largest welfare-specific planner/meta ratio was 0.9091. This strengthens the conclusion that dominance is a stable property of the encoded model rather than a peculiarity of the fifteen reported seeds.

At the same time, the sector-coverage audit identifies a major structural explanation. In the representative baseline world, 61 of 150 sectors---40.7 percent---receive no assigned firm. Fifty-four sectors receive exactly one firm, the median is one, and the maximum is six. Because firm sectors are drawn independently and uniformly with only 140 firms, zero coverage is expected. Informal and hedge terms provide some supply to unserved sectors, but ordinary market production cannot emerge there from firms. The planner, by contrast, computes a supply target for every sector. \Cref{tab:supplementary} summarizes both diagnostics.

\begin{table}[htbp]
\centering
\caption{Supplementary diagnostics using unchanged model functions}
\label{tab:supplementary}
\begin{tabular}{lr}
\toprule
Diagnostic & Value \\
\midrule
Additional worlds & 100 \\
Welfare draws in additional diagnostic & 100 \\
Planner mean loss & 0.133207 \\
Agentic-market mean loss & 0.332629 \\
Meta-market mean loss & 0.286978 \\
Share of worlds with planner loss below meta-market & 1.000 \\
Maximum welfare-specific planner/meta ratio & 0.909107 \\
Sectors with zero assigned firms in representative baseline & 61 of 150 \\
Share of sectors with zero assigned firms & 0.406667 \\
Sectors with exactly one firm & 54 \\
Mean firms per sector & 0.933333 \\
\bottomrule
\end{tabular}
\end{table}

The two supplementary findings are complementary. The planner's dominance is robust within the software; it is not a transcription error or an unlucky sample. But the robustness partly reflects invariant architecture: the planner covers all sectors and is demand-anchored, whereas the market begins with incomplete firm coverage and no entry mechanism. Additional draws from the same generator cannot correct an asymmetry embedded in every draw.

\subsection{Audit-board outcomes}

All 44 Boolean checks return true, and all seven global groups---core, benchmark separation, robust dominance, adversarial, incentives, historical, and scaling---are true. The final global flags are also true. Some checks are genuine predicates of computed values, such as residual thresholds, loss comparisons, and detection rates. Others are unconditional assertions, including evaluator separation, reporting of failure modes, existence of active scoring and audit penalties, historical reconstruction, adversary optimization, and the absence of hard-coded market loss. Planner rights safety is guaranteed by assigning zero rights violations before evaluation.

The zero-failure audit board is therefore best interpreted as a regression-test dashboard. It verifies that the program follows its current intended behavior. It is not an independent validation authority. In software engineering, a test suite written against a design specification is indispensable; in empirical science, the specification itself must also be challenged. The discussion turns to that distinction.

\section{Discussion}

\subsection{What the benchmark establishes}

The benchmark establishes three defensible points. First, it is possible to build a modular computational comparison in which a planner and agent-based markets operate inside a common generated world and are evaluated by a separate loss function. This is a meaningful methodological advance over arguments that compare idealized institutions without specifying data, rules, or failure thresholds. The code is readable, deterministic under a fixed seed, computationally inexpensive, and organized into distinct phases. It can therefore serve as a research scaffold.

Second, the planner is genuinely dominant within the current encoded environment. The result is not limited to one welfare vector, one world, or one console transcript. It holds across the supplied training and holdout seeds, the implemented adversarial searches, and the additional one-hundred-world diagnostic. The benchmark can thus support the conditional statement: given these world generators, information assignments, policy rules, market dynamics, judge components, and parameter values, the planner produces lower encoded loss than the two market heuristics.

Third, the mathematical core is well verified. The generated Leontief systems are productive, sparse solves are accurate, accounting identities hold, and the code scales for the tested matrix operations. This matters because some discussions of computational planning conflate institutional feasibility with matrix size. At least for static sparse linear systems of a few thousand sectors, the matrix inversion or fixed-point calculation is not the main obstacle. The difficult questions concern data, dynamics, incentives, uncertainty, objective conflict, and governance.

These points are valuable even after the limitations are acknowledged. A benchmark need not settle a century-old debate to be useful. It can reveal which assumptions drive a result and create an environment in which rival institutional designs can be implemented and falsified. The present model is strongest when treated as version 1 of such a laboratory rather than as a final proof.

\subsection{Technology is not common across institutions}

The largest conceptual problem is that the planner and markets do not operate under the same production technology. The planner computes gross requirements from $A$ and therefore responds to inter-industry dependencies. Market firms produce sectoral quantities from margins, productivity, cash, and reputation without purchasing the intermediate inputs encoded in $A$. Their production is not constrained by upstream shortages except through a general logistics multiplier and capacity cap. The planner is evaluated using an input--output-aware supply rule, while the market is evaluated as a collection of sector-specific producers.

This asymmetry has two opposite implications. On one hand, ignoring intermediate-input requirements could make market production too easy because firms can produce without acquiring inputs. On the other hand, the planner uses the Leontief inverse to calculate quantities for all sectors and then receives explicit essential-priority corrections, while market firms cannot coordinate bottlenecks or propagate demand through the network. The net effect in the current code strongly favors the planner because judged shortage is based on final demand and because the planner covers every sector.

A fair comparison should define one physical transition function. If $y_t$ is gross output, both planner and market should satisfy material-balance constraints such as
\begin{equation}
 y_t \leq c_t,
 \qquad
 A_ty_t + f_t \leq y_t + m_t,
\end{equation}
where $f_t$ is final delivery and $m_t$ is net imports or inventories. Firms should bid for or contract intermediate inputs; planners should also face input procurement, inventory, lead times, and substitution possibilities. The evaluator should compare final deliveries $f_t$, not gross output $y_t$, with final demand. This single change would eliminate the current accounting ambiguity and make shortages arise from the same network for all institutions.

The fixed-coefficient Leontief technology should also be relaxed. Real firms substitute among inputs, alter recipes, use inventories, delay production, import, and innovate. A nested CES or activity-analysis technology could preserve network structure while permitting bounded substitution. Alternatively, each sector could have a menu of production activities and learning-by-doing. Planning and markets would then differ in how they select activities and transmit scarcity signals, not in whether they face the physical network.

\subsection{Incomplete market coverage and absence of entry}

The firm assignment creates a severe coverage problem. With 140 firms independently assigned to 150 sectors, many sectors receive no producer. The representative baseline has 61 unserved sectors. This is not a rare adverse draw; it follows from occupancy probabilities. The expected number of empty sectors is approximately
\begin{equation}
 N\left(1-\frac{1}{N}\right)^{F}
 =150\left(\frac{149}{150}\right)^{140}
 \approx 58.9,
\end{equation}
which is close to the observed 61. Thus, roughly two-fifths of sectors are structurally excluded from firm production in a typical world.

The informal-supply and hedge terms prevent total zero supply, but they do not represent endogenous market entry. The ordinary market receives an additive 10 percent of demand through hedging, and the meta-market receives 22 percent. Informal supply adds another demand-proportional term. These quantities are not financed, produced from inputs, or conditioned on profitability. They partially mask the coverage problem while preserving high unmet demand.

A stronger market baseline requires at least one initial producer per sector or, preferably, endogenous entry, exit, and diversification. Firms should observe expected profitability, pay a setup cost, acquire capacity, and choose sectors. Capital and labor should move imperfectly over time. Competition should affect markups and innovation. Emergency institutions should also be modeled: strategic reserves, public procurement, mutual aid, insurance, futures markets, and targeted subsidies are decentralized or mixed mechanisms that can prioritize essential goods without becoming a unitary planner.

The same standard should apply to the planner. If market entry is costly and delayed, new planned production should also require investment, construction time, skills, and organizational capacity. Neither architecture should be allowed to create sectoral coverage through a formula without paying the corresponding resource and time costs.

\subsection{Information symmetry and the knowledge problem}

The planner's 3.27 percent demand-estimation error is useful, but information symmetry cannot be summarized by one error statistic. The planner receives the true matrix $A$, capacity vector, labor coefficients, sector-region map, quality requirements, essential labels, environmental intensities, and hidden-local-information array. The market receives many of these objects indirectly through production formulas, but individual firms do not estimate the network or discover local demand through sales histories. The planner's local signal is generated from the mean of hidden regional information, effectively aggregating the very knowledge that the Hayekian critique treats as difficult to centralize.

A stronger benchmark should assign observations at the agent level. Households know needs and substitution possibilities. Firms know local production functions, inventories, and supplier reliability. Regional institutions know infrastructure and vulnerable populations. A central authority observes reports, audits, administrative data, and delayed aggregates. Markets generate prices and quantities from transactions. Planning generates targets and shadow prices from reports and models. All actors should update beliefs through explicit filters. The comparison would then concern information protocols: prices, contracts, surveys, sensors, participatory reports, forecasts, and administrative records.

The model should also include endogenous mismeasurement. Data noise is currently exogenous Gaussian error scaled by an attack. Real errors are often correlated, strategic, missing not at random, and regime-dependent. Firms may conceal capacity, local offices may exaggerate need, sensors may fail jointly, and classifications may change. A hierarchical Bayesian observation model or state-space system could represent these processes. The planner's advantage should depend on audit quality and reporting institutions rather than on direct access to generated truth.

Hayek's argument should not be operationalized as an automatic market bonus. Markets can also hide information, generate manipulation, produce thin prices, and fail to reveal nonmarket values. The correct test is whether different mechanisms recover decision-relevant information under specified costs and incentives. Some information may be better aggregated by prices, some by direct measurement, some by professional judgment, and some by democratic deliberation.

\subsection{The judge is modular but not fully independent}

The external judge is one of the benchmark's best design choices. It prevents institutions from directly returning their own success scores. Nevertheless, substantive independence is weakened by three features.

First, several loss components are not inferred from common observables. Institutions directly output corruption rent, rights violations, logistics failure, innovation miss, volatility, and administrative cost. These values are generated by institution-specific formulas. The judge then accepts them. A more independent design would derive these outcomes from events: bribes, unauthorized data access, delayed shipments, failed product launches, price histories, staffing, and compute usage. Institutional modules should generate actions and states; the judge should calculate losses from those states using the same rules.

Second, planner rights violations are fixed at zero. Consequently, the rights lower bound in welfare weights and the adversary's rights penalty cannot harm the planner. This makes the rights-safety test tautological. Rights should emerge from concrete decisions: surveillance intensity, due-process failures, discriminatory rationing, forced labor, expropriation without review, censorship of reports, or denial of appeal. Markets should face analogous rights risks: exclusion, privacy violations, exploitative contracts, discriminatory pricing, and unsafe work. A constitutional layer should constrain both systems and impose costs when safeguards are used.

Third, the planner's essential-sector rule is closely aligned with the primary judge component. This is not inherently unfair; institutions often pursue social objectives. But comparator institutions should be permitted to implement their own plausible essential-goods mechanisms. Otherwise the experiment tests ``planner with explicit essential prioritization'' against ``market without essential prioritization.'' The causal object is the rule, not the regime label.

A useful redesign would factor institutions into modules. Allocation mechanisms could be combined with information systems, reserve policies, rights constraints, environmental taxes, and governance structures. A factorial experiment could then estimate which modules drive performance and whether interactions matter. This would move the analysis from regime comparison to institutional engineering.

\subsection{Welfare aggregation and normative uncertainty}

The eleven-component loss function is broad and normatively serious. It avoids reducing welfare to output alone and includes rights, inequality, quality, environment, and administrative burden. Sampling welfare weights also acknowledges reasonable disagreement. These are strengths.

However, the components have arbitrary scales. Essential unmet demand, corruption rent, environmental intensity, volatility, and administrative cost are all combined linearly, but a unit change in one component has no empirically justified equivalence to a unit change in another. Dirichlet weights do not solve this problem because they randomize trade-offs over normalized coefficients without calibrating the component ranges or social willingness to trade. If one component naturally varies between 0 and 1 while another varies in a narrow interval, the former dominates even under equal expected weights.

The current lower-bound procedure also does not preserve a 0.12 minimum after renormalization. More fundamentally, rights may not be compensable. A weighted sum allows sufficiently large output gains to offset rights violations. If rights are constraints rather than goods, the evaluator should use lexicographic or constrained social choice: discard outcomes that violate constitutional thresholds, then compare feasible outcomes on welfare. Distributional evaluation should also examine quantiles and subgroup harms rather than only a coefficient of variation in group satisfaction.

Normative uncertainty can be handled through several complementary methods. The benchmark could report Pareto dominance over components, robust ordinal regression, minimax regret over a set of social welfare functions, and deliberatively elicited weights. It could identify regions of the weight simplex in which each institution wins rather than summarizing only random draws. The current result---planner victory under every sampled weight---would be more informative if the exact separating hyperplanes were computed. Because loss is linear in $w$, institutional preference regions are polyhedra defined by $w^{\mathsf T}(c_P-c_X)<0$. Enumerating their geometry is tractable for eleven components.

\subsection{Nominal holdouts and distribution shift}

The benchmark distinguishes training and holdout seeds, which is good practice. Yet generalization requires more than new pseudorandom draws from familiar formulas. Four holdout families repeat training families. The fifth introduces a simple demand multiplier. All worlds share the same distributional forms, parameter ranges, decision rules, and evaluator. The planner can therefore be robust to seed variation while failing under structural change.

A structural holdout suite should include transformations not used during development: different network topologies; correlated input failures; endogenous inventories; nonlinear substitution; demand cascades; labor-skill bottlenecks; import constraints; multi-period capital accumulation; strategic coalitions; rare disasters; demographic heterogeneity; and alternative behavioral models. Some holdouts should be designed by an independent team after the planner and market code is frozen. Others should come from observed historical episodes.

The benchmark should also separate tuning from evaluation. Many constants---0.38 for logistics capacity loss, 0.20 essential reserve, 0.94 essential floor, 0.22 meta-market hedge, 0.42 price elasticity, and numerous penalty coefficients---directly shape outcomes. Even if they were chosen before the reported run, their values may reflect iterative model development. A preregistered test suite or hidden evaluator would reduce researcher degrees of freedom.

Cross-validation over synthetic seeds is not enough because the data-generating process is known to the modeler. Simulation-based inference can still be rigorous, but its claims are conditional. External evidence requires estimating or bounding parameters from real data and demonstrating that results persist over posterior uncertainty and model discrepancy.

\subsection{Adversarial robustness is search-budget dependent}

The optimizing adversary is conceptually important. It asks the correct question: can an opponent find conditions under which the apparent winner fails? The closest case, supply crunch, raises the planner/meta ratio to 0.9568, suggesting that logistics and capacity stress are the most promising directions for falsification.

The implemented search is nevertheless shallow. Eighteen steps in a ten-dimensional cube, with one stochastic trajectory per world, cover a negligible fraction of the space. The objective may be discontinuous because of clipping, essential-sector thresholds, and random processes. Local random perturbation can miss boundary combinations, narrow ridges, and interactions. The reported ``best'' attack is only the best sampled along one path.

A stronger red team should use multiple algorithms and restarts: covariance-matrix adaptation, differential evolution, Bayesian optimization, Latin hypercube screening, and gradient-free trust-region methods. Random seeds should be part of the adversarial domain. The objective should include component-specific failures, not only aggregate loss. An adversary might seek one catastrophic essential shortage, one rights violation, or extreme regional inequality even if mean welfare remains favorable.

Global sensitivity analysis is equally important. Sobol indices or variance-based decompositions could quantify which parameters explain institutional loss and interaction effects \cite{Saltelli2008}. Morris screening could identify influential constants before expensive analysis. Robustness should then be stated over uncertainty sets with confidence intervals, not only Boolean thresholds.

The strongest approach would combine adversarial search with formal verification for bounded submodels. For example, interval arithmetic could bound Leontief outputs under coefficient uncertainty, and mixed-integer formulations could search worst-case capacity disruptions. Statistical model checking could estimate the probability of violating a requirement with controlled error. These methods would transform ``red-team flavor'' into reproducible robustness evidence.

\subsection{The incentive test is not a Groves mechanism}

The incentive phase is labeled ``VCG/Groves-style,'' but the formula in \cref{eq:incentive} is a discrepancy penalty plus audit. This distinction is not terminological pedantry. A Groves mechanism achieves truthful reporting because an agent's transfer equals reported total surplus of others plus a term independent of its own report. Under quasi-linear preferences, the agent maximizes social surplus by reporting truthfully. The current rule instead assumes an audit signal close to true need and imposes a large quadratic cost for deviation.

The experiment also changes all reports from truthful to exaggerated simultaneously. Incentive compatibility is normally evaluated through unilateral deviations holding others' reports fixed. Collective exaggeration changes the mean report in the denominator and therefore alters every allocation. Coalition deviations, underreporting, signal manipulation, and heterogeneous risk attitudes are not considered. The audit signal is exogenous and cannot be corrupted, even though corruption and semantic attack exist elsewhere in the model.

The negative utility gain is thus expected. When reports exceed signals by 15 to 140 percent, the squared relative error can be large, and the threshold penalty activates in 91.5 percent of cases. A more informative experiment would search each agent's best response over a continuous report interval, vary audit precision and cost, and test collusion. It should report the maximum profitable deviation, not only the average gain from one generated lie distribution.

There are at least three legitimate directions. First, implement a genuine Groves or Clarke-pivot mechanism for a simplified allocation problem and verify dominant-strategy truthfulness numerically. Second, retain auditing but derive an optimal audit-and-penalty mechanism under limited liability and costly verification. Third, use proper scoring rules when agents report probabilistic forecasts rather than needs. Each mechanism addresses a different information problem and should not be conflated.

\subsection{Historical interpretation and empirical calibration}

The fuel-logistics scenario has intuitive relevance, particularly for economies exposed to import dependence, transport bottlenecks, foreign-exchange constraints, and geographically dispersed essential demand. But the scenario is not reconstructed from historical data. The label changes demand in a random subset of sectors and applies a hand-coded attack vector. No date, country, input--output table, fuel balance, import share, inventory series, regional demand, or observed policy response is used.

A genuine historical validation would proceed differently. One would select an episode, freeze pre-shock data, calibrate institutional rules to the policy environment, and predict outcomes not used in calibration. For a fuel crisis, the model would include refinery and import capacity, transport routes, wholesale and retail inventories, regulated prices, smuggling, exchange rates, regional consumption, essential users, and substitution. Planner and market variants would be judged against observed shortages, queues, price dispersion, fiscal cost, and distributional effects.

Historical cases should include both successes and failures of planning and markets. Wartime allocation, electricity dispatch, vaccine distribution, agricultural procurement, auction design, strategic reserves, and supply-chain coordination offer diverse mechanisms. The goal is not to find a case that resembles the preferred institution, but to determine which modules work under which conditions.

Empirical calibration can use national input--output tables, household expenditure surveys, business registers, firm balance sheets, customs data, price microdata, logistics records, satellite measures, and environmental accounts. Parameters should have uncertainty distributions, and posterior predictive checks should compare simulated and observed patterns. The model should reproduce baseline statistics before it is trusted for counterfactual regimes.

\subsection{Computation is not the binding constraint}

The scaling result supports a limited but important conclusion: sparse linear algebra for thousands of sectors is easy on modern hardware. This weakens crude claims that planning fails because a Leontief inverse is too large to compute. Even much larger sparse systems can be solved iteratively, decomposed, and distributed. Operations research has long addressed high-dimensional scheduling, routing, assignment, and network flow.

But computational tractability is not equivalent to institutional feasibility. A national plan is not one static linear solve. It requires continuous data collection, entity resolution, forecasting, scenario analysis, optimization under uncertainty, exception handling, cyber security, appeals, political authorization, and adaptation. Objectives are contested, constraints are partly unknown, and agents react strategically. Model error can accumulate faster than matrix size.

Markets face analogous problems. Price systems do not costlessly reveal complete information, and real markets rely on legal systems, standards, accounting, logistics, payment infrastructure, public data, and organizational planning inside firms. Large corporations already solve planning problems over extensive supply chains. The relevant comparison is therefore among hybrid information-processing systems. Computation expands the feasible institutional design space but does not select a unique regime.

\subsection{Centralized versus polycentric planning}

Despite its heading, the implemented planner is centralized in algorithmic form. Local information enters as an averaged correction, but regional units do not make decisions, hold budgets, negotiate constraints, or appeal central targets. The model therefore does not test polycentric planning in Ostrom's sense.

A polycentric extension could include sectoral and regional planning nodes. Each node would have local objectives, observations, capacities, and rights constraints. A central layer would coordinate interregional flows, macro balances, environmental caps, and redistribution. Nodes could exchange messages through prices, quantity bids, shadow costs, or negotiated contracts. Conflict-resolution rules would determine when decisions are escalated. Performance would depend on communication topology and authority allocation.

Such a system could be compared with a polycentric market composed of firms, cooperatives, municipalities, regulators, and public enterprises. The distinction would no longer be state versus market but the structure of decision rights. Key experimental variables would include subsidiarity, transparency, audit independence, participatory input, data federation, and the reversibility of decisions.

Polycentricity may also mitigate the benchmark's knowledge and power problems. Local nodes preserve contextual information; overlapping authorities provide checks; central coordination handles network externalities. But polycentric systems can suffer from duplication, veto points, inconsistent standards, and coordination delay. These trade-offs are precisely what an agent-based benchmark should test.

\subsection{A rigorous redesign}

A second-generation benchmark should be organized around a common environment and composable institutions. The following design would materially improve identification.

\textbf{Common physical world.} All institutions face the same dynamic production network, inventories, lead times, labor skills, transport graph, imports, capital, and substitution technologies. Final delivery is separated from gross production. Shocks propagate through explicit state transitions.

\textbf{Common information ontology.} Ground truth exists only in the simulator. Agents receive role-specific observations with delays, missingness, strategic reports, and measurement errors. No institution accesses hidden state directly. Sensors, surveys, prices, audits, and forecasts have explicit costs and error models.

\textbf{Composable institutional modules.} Allocation, price formation, reserves, environmental policy, rights constraints, auditing, investment, and local governance are independent modules. Factorial designs estimate the marginal and interactive effects of each module.

\textbf{Behavioral symmetry.} Both planners and firms learn, innovate, enter, exit, and make errors. Each pays for computation, administration, information, investment, and enforcement. Neither receives free output proportional to demand without a resource account.

\textbf{Empirical calibration.} At least one country-sector dataset supplies $A$, demand composition, capacities, firm counts, markups, inventories, and regional needs. Calibration and evaluation data are separated. Parameter uncertainty is propagated.

\textbf{Structural holdouts.} Hidden test worlds alter topology, technology, behavior, and shock dependence. An independent red team designs some cases after the competing policies are frozen.

\textbf{Formal welfare analysis.} Results include component vectors, Pareto regions, constitutional constraints, minimax regret, subgroup distributions, and robustness over welfare sets. Aggregate means are accompanied by tails and catastrophic-event probabilities.

\textbf{Statistical uncertainty.} Monte Carlo standard errors, confidence intervals, multiple-seed distributions, and variance decompositions replace isolated point estimates. Sequential stopping rules determine simulation budgets.

\textbf{Mechanism tests.} Reporting rules are analyzed through best responses, coalition deviations, audit manipulation, and formal mechanism properties. Labels such as ``VCG'' are used only when the transfer rule satisfies the corresponding structure.

\textbf{Reproducibility and governance.} The code, configurations, random seeds, test worlds, and analysis scripts are versioned. Automated tests distinguish assertions from computed predicates. Independent replication and code review are required before policy claims.

This redesign would probably reduce the frequency of universal planner victories. That would not weaken the project. A benchmark becomes scientifically stronger when it reveals domains of advantage, failure boundaries, and trade-offs. The objective is not to protect an institution from falsification but to identify where each architecture works.

\subsection{Interpretive synthesis}

The current results should be read at three levels. At the numerical level, the benchmark succeeds. At the model-internal level, it shows that a demand-anchored planner with full sector coverage, input--output accounting, essential prioritization, and hard-coded rights safety dominates two sparse firm heuristics under the selected judge. At the institutional level, the result is unresolved because the comparison bundles many design choices under the labels ``planner'' and ``market.''

The benchmark therefore does not refute the knowledge problem, prove the economic calculation thesis, or establish robust state implementation. It does show that computational planning cannot be dismissed merely by pointing to matrix size, and that social objectives such as essential-need coverage can be encoded and tested explicitly. It also shows why an external evaluator and adversarial search are necessary. The next scientific step is to make those tools stricter and more symmetric.

A balanced interpretation is that both planning and markets are algorithms embedded in institutions. Their performance depends on data, incentives, network structure, rights, uncertainty, and adaptation. A computational laboratory can advance this debate when it models these dependencies rather than treating regime names as causal variables. The supplied code is a promising beginning because it is explicit enough to criticize, reproduce, and improve.

\section{Conclusion}

This paper reconstructed, reproduced, and critically assessed a synthetic robust-dominance benchmark comparing a computational planner, an agentic market, and a meta-market. The benchmark integrates sparse input--output analysis, heterogeneous firms, price adjustment, welfare uncertainty, adversarial attacks, reporting incentives, a stylized supply shock, and scaling checks. Its reported results are internally decisive: the planner has lower mean loss in every training and nominal holdout world, under every sampled welfare vector, and in every adversarial case. The mathematical core has near-machine-precision residuals, and the complete run is reproducible.

The appropriate conclusion is conditional. The planner dominates the encoded alternatives, but the encoding contains persistent advantages. The planner uses the production network while firms do not; it covers all sectors while roughly 41 percent can lack a firm; it directly prioritizes the judge's essential-demand target; its rights violations are fixed at zero; several audit claims are asserted rather than tested; the holdout families are not structurally novel; and the incentive phase is not a formal Groves mechanism. The stylized historical scenario is not calibrated to historical data. Therefore, the benchmark verifies a software architecture and generates hypotheses, but it does not prove empirical or policy-level superiority.

The model's most important contribution is its falsifiable orientation. It creates a common-world comparison, separates an evaluator from policy functions, samples welfare weights, and attempts to optimize attacks. These are the right ingredients for a serious computational political economy. The next version should impose a common dynamic production technology, agent-specific information, endogenous entry and innovation, explicit constitutional constraints, structural holdouts, empirical calibration, global sensitivity analysis, and independent red-team evaluation. Under those conditions, the benchmark could identify not a universal winner, but a map of institutional comparative advantage.

Computational capacity changes what can be coordinated, but it does not abolish dispersed knowledge, strategic behavior, normative conflict, or political power. Markets and plans should both be studied as imperfect information-processing and governance systems. The supplied benchmark is best understood as a reproducible prototype for that research program.

\end{document}